%% file: e08014.tex
\documentclass[aps,prl,superscriptaddress,showpacs,twocolumn,floatfix,amsmath,amssymb]{revtex4-1}

\usepackage{graphicx} 
\usepackage{hyperref} 
\usepackage[mathlines]{lineno} 
\def\gtorder{\mathrel{\raise.3ex\hbox{$>$}\mkern-14mu
\lower0.6ex\hbox{$\sim$}}}
\def\ltorder{\mathrel{\raise.3ex\hbox{$<$}\mkern-14mu
\lower0.6ex\hbox{$\sim$}}}

\begin{document}

\title{Search for three-nucleon short-range correlations in light nuclei}

\input{authors.tex}


\date{\today}

\begin{abstract}

We present new data probing short-range correlations (SRCs) in nuclei through the measurement of electron scattering
off high-momentum nucleons in nuclei. The inclusive $^4$He/$^3$He cross section ratio is observed to be both $x$ and
$Q^2$ independent for $1.5 < x <2$, confirming the dominance of two-nucleon short-range correlations. For $x>2$, our
data support the hypothesis that a previous claim of three-nucleon correlation dominance was an artifact caused by
the limited resolution of the measurement. While 3N-SRCs appear to have an important contribution, our data show
that isolating 3N-SRCs is significantly more complicated than for 2N-SRCs.

\end{abstract}

\maketitle

\section{INTRODUCTION}
Understanding the complex structure of the nucleus remains one of the major uncompleted tasks in nuclear physics, and the
high-momentum components of the nuclear wave-function continue to attract attention~\cite{arrington12, fomin17, hen17}.
Momenta above the Fermi momentum are strongly suppressed in shell model and mean field calculations~\cite{DeForest1983}.
Subsequently, these calculations under-predict (over-predict) the cross section for proton knock-out reactions
above (below) the Fermi momentum~\cite{VanDerSteenhoven1988547, Lapikas1993297, Kelly:1996hd}.

In the dense and energetic environment of the nucleus, nucleons have a significant probability of
interacting at distances $\le$1~fm, even in light nuclei~\cite{carlson14,lu13}. Protons and neutrons
interacting through the strong, short-distance part of the NN interaction give rise to pairs of
nucleons with large momenta. These high-momentum pairs, referred to as short-range correlations (SRCs), 
generate high-momentum nucleons in nuclei~\cite{Frankfurt1981215, SLAC_Measurement_PRC.48.2451, src_john}.
These are the primary source of nucleons above the Fermi momentum, $k_F \approx 250$~MeV/$c$. For momenta below $k_F$, we observe shell-model behavior which is
strongly $A$ dependent, while two-body physics dominates above $k_F$ resulting in a universal
structure for all nuclei that is driven by the details of the NN interaction~\cite{RevModPhys.80.189,
PhysRevC.53.1689, wiringa14}.

In the case of inclusive electron-nucleus scattering, it is possible to select kinematics that isolate scattering from high-momentum
nucleons. The electron transfers energy, $\nu$, and momentum, $\vec{q}$,
to the struck nucleon by exchanging a virtual photon with four momentum transfer $q^2=-Q^2=\nu^2-|\vec{q}|^2$.
It is useful to define the kinematic variable $x = Q^2/(2M_p\nu)$, where $M_p$
is the mass of the proton. Elastic scattering from a stationary proton corresponds to $x=1$,
while inelastic scattering must occur at $x<1$ and scattering at $x>1$ is kinematically forbidden.
In a nucleus, the momentum of the nucleon produces a broadened quasielastic peak centered near $x=1$.
At values of $x$ slightly greater than unity, scattering can occur from either low-momentum nucleons
or from the high-momentum nucleons associated with SRCs. As $x$
increases, larger initial momenta are required until scattering from nucleons below the Fermi
momentum is kinematically forbidden, isolating scattering from high-momentum nucleons associated with
SRCs~\cite{RevModPhys.80.189, PhysRevC.53.1689, src_john, egiyan2003}.

Because the momentum distribution of the nucleus is not a physical observable, one cannot directly extract
and study its high-momentum component. One can, however, test the idea of a universal structure at
high-momenta by comparing scattering from different nuclei at kinematics which require that the
struck nucleon have a high initial momentum~\cite{Frankfurt1981215, SLAC_Measurement_PRC.48.2451}.
Several experiments at SLAC and Jefferson Lab studied inclusive scattering at $x>1$ to compare scattering from high-momentum nucleons in light and heavy nuclei~\cite{SLAC_Measurement_PRC.48.2451, egiyan2003, PhysRevLett.96.082501, fomin2012, src_john,
arrington99, arrington01}. These measurements confirmed the picture of a universal form to the scattering in the region dominated by high-momentum nucleons
The cross section ratios for inclusive scattering from
heavy nuclei to the deuteron were shown to scale, i.e. be independent of $x$ and $Q^2$, for $x \gtorder 1.5$
and $Q^2 \gtorder 1.5$~GeV$^2$, corresponding to scattering from nucleons with momenta above 300 MeV/$c$.
Other measurements have demonstrated that these high-momentum components are dominated by high-momentum $np$
pairs~\cite{aclander99, tang03, Subedi13062008, korover2014, hen14_science, piasetzky06}, meaning that the
high-momentum
components in all nuclei have a predominantly deuteron-like structure. While final-state interactions (FSI)
decrease with increasing $Q^2$ in inclusive scattering, FSI between nucleons in the correlated pair may not
disappear. It is typically assumed that the FSI are identical for the deuteron and the deuteron-like pair in
heavier nuclei, and thus cancel in these ratios~\cite{Frankfurt1981215, src_john}, although this is not true
for all attempts to calculate FSI effects~\cite{RevModPhys.80.189}.

This approach can be extended to look for universal behavior arising from 3N-SRCs by examining scattering at
$x>2$ (beyond the kinematic limit for scattering from a deuteron). Within the simple SRC
model~\cite{Frankfurt1981215}, the cross section is composed of scattering from one-body, two-body,
etc... configurations, with the one-body (shell-model) contributions dominating at $x \approx 1$, while
2N(3N)-SRCs dominate as $x \to 2(3)$. Taking ratios of heavier nuclei to $^3$He allows a similar
examination of the target ratios for $x>2$, where the simple SRC model predicts a universal behavior
associated with three-nucleon SRCs (3N-SRCs) - configurations where three nucleons have large relative
momenta but little total momentum. 3N-SRCs could come from either three-nucleon forces or multiple hard
two-nucleon interactions. The first such measurement~\cite{PhysRevLett.96.082501} observed $x$-independent
ratios for $x > 2.25$. This was interpreted as a result of 3N-SRCs dominance in this region.
However, the ratios were extracted at relatively small $Q^2$ values and the $Q^2$ dependence was not measured.
In the experiment of Ref.~\cite{fomin2012}, at higher $Q^2$, the $^4$He/$^3$He ratios  were significantly
larger. Consequently, the question of whether 3N-SRC contributions have been cleanly identified and observed
to dominate at some large momentum scale is as yet unanswered.

\section{EXPERIMENTAL DETAILS}
The results reported here are from JLab experiment E08-014~\cite{e08014_pr}, which focused on precise
measurements of the $x$ and $Q^2$ dependence of the $^4$He/$^3$He cross section ratios at large $x$. A
$3.356$~GeV electron beam with currents ranging from 40 to 120 $\mu$A impinged on nuclear targets, and
scattered electrons were detected in two nearly identical High-Resolution Spectrometers (HRSs)~\cite{halla_nim}.
Data were taken on three 20-cm cryogenic targets (liquid $^2$H and gaseous $^3$He and $^4$He) and on
thin foils of $\mathrm{^{12}C}$ and $\mathrm{^{40,48}Ca}$.

Each HRS consists of a pair of vertical drift chambers (VDCs) for particle tracking, two scintillator planes
for triggering and timing measurements, and a gas \v{C}erenkov counter and two layers of lead-glass
calorimeters for particle identification~\cite{halla_nim}. Scattering was measured at $\theta=21^\circ$,
$23^\circ$, $25^\circ$, and $28^\circ$, covering a $Q^2$ range of 1.3--2.2~GeV$^2$. A detailed description
of the experiment and data analysis can be found in Ref.~\cite{zye_thesis}.


The data analysis is relatively straightforward, as inclusive scattering at $x>1$ yields modest rates and a
small pion background. The trigger and tracking inefficiencies are small and applied as a correction to the
measured yield. Electrons are identified by applying cuts on the signals from both the \v{C}erenkov detector
and the calorimeters. The cuts give $>99$\% electron efficiency with negligible pion contamination. The
overall dead-time of the data acquisition system (DAQ) was evaluated on a run-by-run basis. To ensure a
well-understood acceptance, the solid angle and momentum were limited to high-acceptance regions
and a model of the HRSs was used to apply residual corrections~\cite{zye_thesis}.

The scattered electron momentum, in-plane and out-of-plane angles, and vertex position at the target can be
reconstructed from the VDC tracking information. The transformation from focal plane to target quantities has
been obtained from previous experiments, but for the right HRS, the third quadrupole could not achieve its
nominal operating current, so data were taken with a 15\% reduction in its field. New optics data were taken
to correct for the modified tune. Many of the systematic uncertainties in the spectrometers are correlated,
and so when merging data from the two spectrometers, we add the statistics and then apply
the systematic uncertainties to the combined result. 


The $^{3,4}$He targets have a large background from scattering in the cell walls. We apply a $\pm 7$~cm
cut around the center of the target, removing $>99.9\%$ of the events from target endcap
scattering, as determined from measurements on empty target cells. One of the largest contributions to the
systematic uncertainty comes from target density reduction due to heating of the $^2$H, $^3$He, and $^4$He
targets by the high-current electron beam. We made dedicated measurements over a range of beam currents
and used the variation of the yield to determine the current dependence of the target density. We observed
a large effect that varied with the position along the target, and the extrapolation to zero current
did not yield a uniform density. This indicates a non-linear current dependence that is not uniform over
the length of the target, making it difficult to determine the absolute target thickness.
However, the size of the effect is similar for $^3$He and $^4$He, and the $^4$He/$^3$He ratios are consistent
with previous data near the quasielastic peak and in the 2N-SRC region. We therefore assume that the error in
extrapolating to zero current largely cancels in the ratio and apply a 5\% scale uncertainty for the
$^4$He/$^3$He ratios. For the absolute uncertainty, the $^{3,4}$He targets have a large normalization
uncertainty, potentially 10\% or larger. This does not impact our study of 3N-SRCs,
and so we do not attempt to normalize the data to existing measurements.

The measured yields, corrected for inefficiencies and normalized to the integrated luminosity, were binned
in $x$ and compared to the simulated yield. The simulation uses a $y$-scaling cross section
model~\cite{day_arns, arrington99} with radiative corrections applied using the peaking
approximation~\cite{stein75}. Coulomb corrections are applied within an improved effective
momentum approximation~\cite{aste05,arrington12}, and are 2\% or smaller for all data presented here. The
uncertainty in the target thicknesses dominates the total scale uncertainty (5.1\%) of the ratios, while
density fluctuations and dummy subtraction (used to remove the contribution from the Aluminum endcaps of the target) dominate the point-to-point systematic uncertainty of 1.3\%.

\section{RESULTS} 

\begin{figure}[!ht]
    \begin{center}
        \includegraphics[width=8.6cm,angle=0]{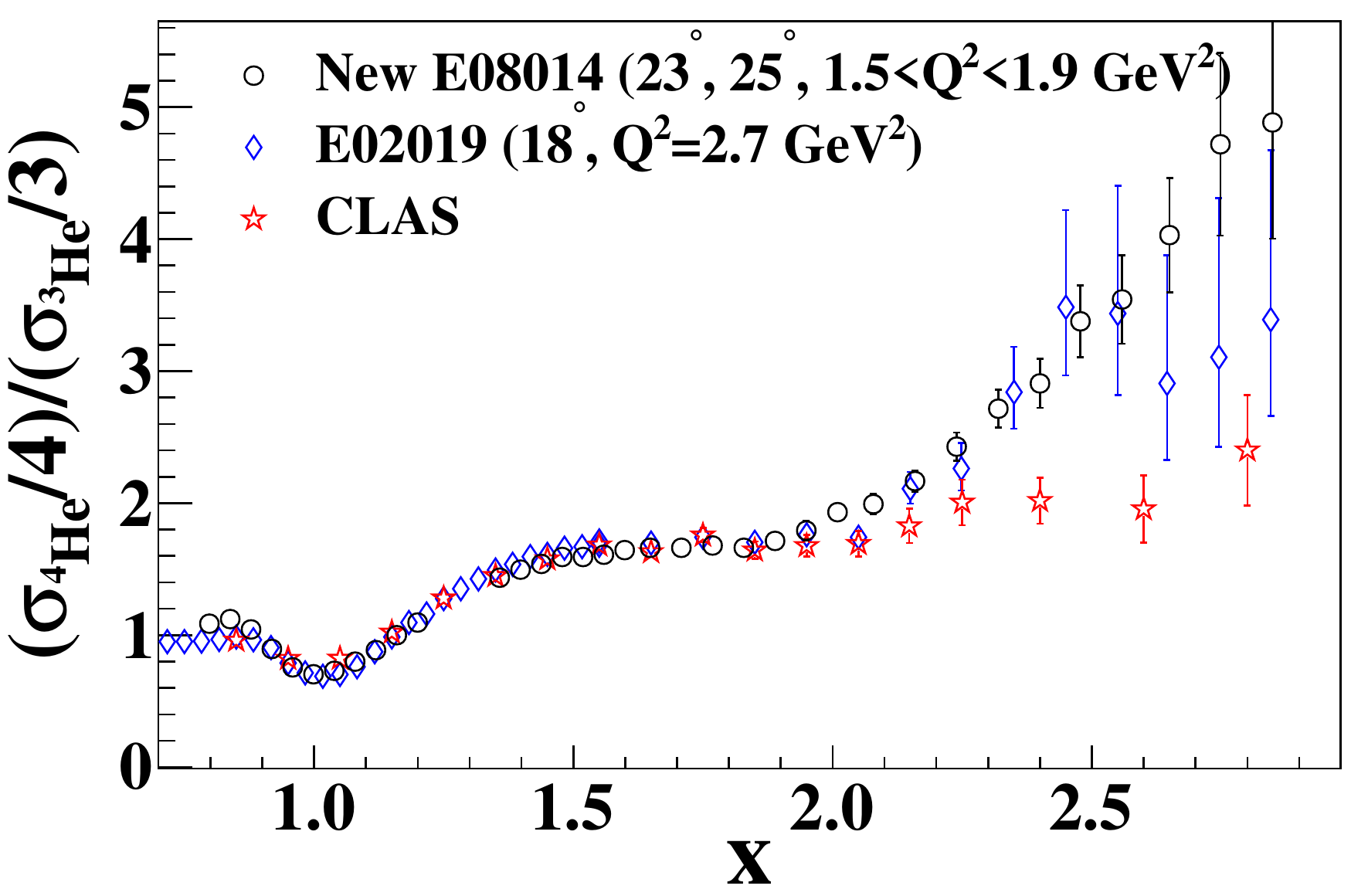}
    \end{center}
    \caption{(Color online) The $^4$He/$^3$He cross section ratio for $Q^2>1.4$~GeV$^2$, along with results
        from CLAS~\cite{PhysRevLett.96.082501} and Hall C (E02-019)~\cite{fomin2012}. The error bars include
        statistical and systematic uncertainties; the 5.1\% scale uncertainty is not shown.}
    \label{fig:ratios_highqsq}
\end{figure}

Figure~\ref{fig:ratios_highqsq} presents the $^4$He/$^3$He cross section ratio for measurements with
$Q^2 > 1.4$~GeV$^2$, obtained by combining the ratios from 23$^\circ$ and 25$^\circ$ scattering. In the 2N-SRC region,
our data are in good agreement with the CLAS~\cite{PhysRevLett.96.082501} and
Hall C~\cite{fomin2012} results, revealing a plateau for $1.5 < x < 2$. At $x>2$, our
ratios are significantly larger than the CLAS data, but consistent with the Hall C
results. This supports the explanation provided in a recent comment~\cite{Higinbotham:2014xna}
which concluded that the observed plateau was likely the result of large bin-migration effects resulting from
the limited CLAS momentum resolution.

While the rise in the ratio above $x=2$ indicates contributions beyond 2N-SRCs, we do not observe the 3N-SRC
plateau expected in the naive SRC model~\cite{Frankfurt1981215,SLAC_Measurement_PRC.48.2451}. In this model,
the prediction of scaling as an indication of SRC dominance is a simple and robust way to test for 2N-SRCs.
It is much less clear how well it can indicate the
presence of 3N-SRCs. For 2N-SRCs, one can predict \textit{a priori} where the plateau should be observed:
for a given $Q^2$ value, $x$ can be chosen to require a minimum nucleon momentum above the Fermi
momentum, strongly suppressing single-particle contributions. It is not clear what values of $x$ and $Q^2$
are required to suppress 2N-SRC contributions well enough to isolate 3N-SRCs. Much larger $Q^2$ values may
be required to isolate 3N-SRCs and see analogous plateaus at $x>2.5$~\cite{fomin17}.

\begin{figure}[!ht]
    \begin{center}
        \includegraphics[width=8.6cm,angle=0]{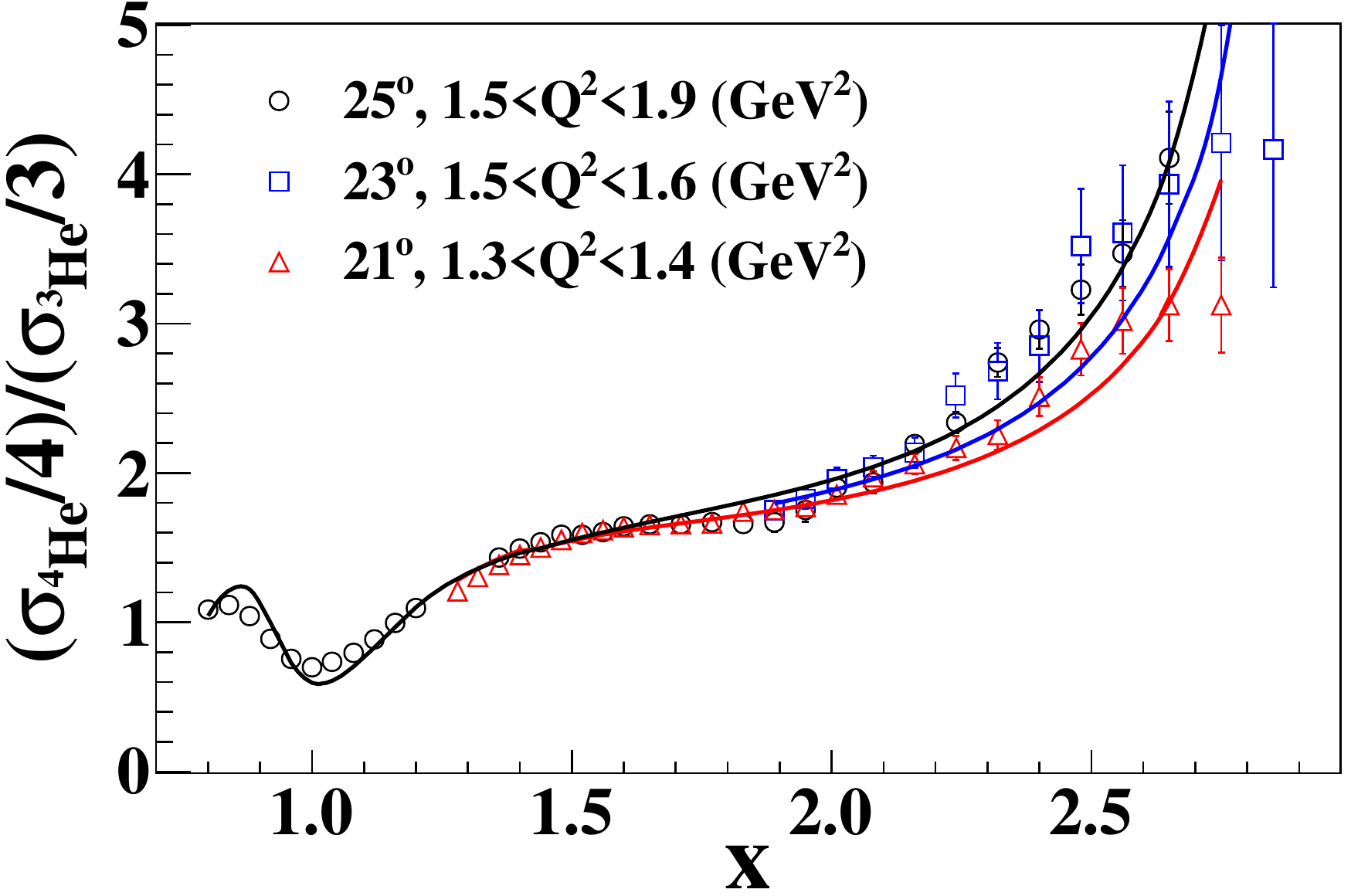}
    \end{center}
    \caption{(Color online) The $^4$He/$^3$He (top) cross section ratios for all angles. The solid lines
    are ratios from our $y$-scaling cross section model based on a parameterized longitudinal momentum
    distribution $F(y)$. The 5.1\% normalization uncertainty is not shown.}
    \label{fig:ratios_allqsq}
\end{figure}

For A/$^2$H ratios, the plateau must eventually disappear as the deuteron cross section falls to zero for $ x \to
M_D / M_p\approx 2$, causing the ratio to rise sharply to infinity. Both the previous high-$Q^2$ deuterium data and
our simple cross section model, based on a parameterization of the longitudinal momentum distribution, show that the
sharp drop of the deuteron cross section does not occur until $x \approx 1.9$, resulting in a clear plateau
for $1.5 < x < 1.9$. For $^3$He, our model shows a similar falloff of the $^3$He cross section
starting near $x \approx 2.5$, producing a rise in the A/$^3$He ratio that sets in well below the kinematic limit
$x \approx 3$. This rapid rise in the A/$^3$He ratio as one approaches the $^3$He kinematic threshold shifts to
lower $x$ as $Q^2$ increases, as seen in both the data and model in Fig~\ref{fig:ratios_allqsq}. So while
the plateau is expected to set in at lower $x$ values as $Q^2$ increases, as seen in the 2N-SRC
region~\cite{ SLAC_Measurement_PRC.48.2451, PhysRevLett.96.082501}, the large-$x$ breakdown also shifts to
lower $x$ values, potentially limiting the $x$ range over which a plateau could be observed, even in the case
of 3N-SRC dominance.

\begin{figure}[!ht]
    \begin{center}
        \includegraphics[width=8.6cm,angle=0]{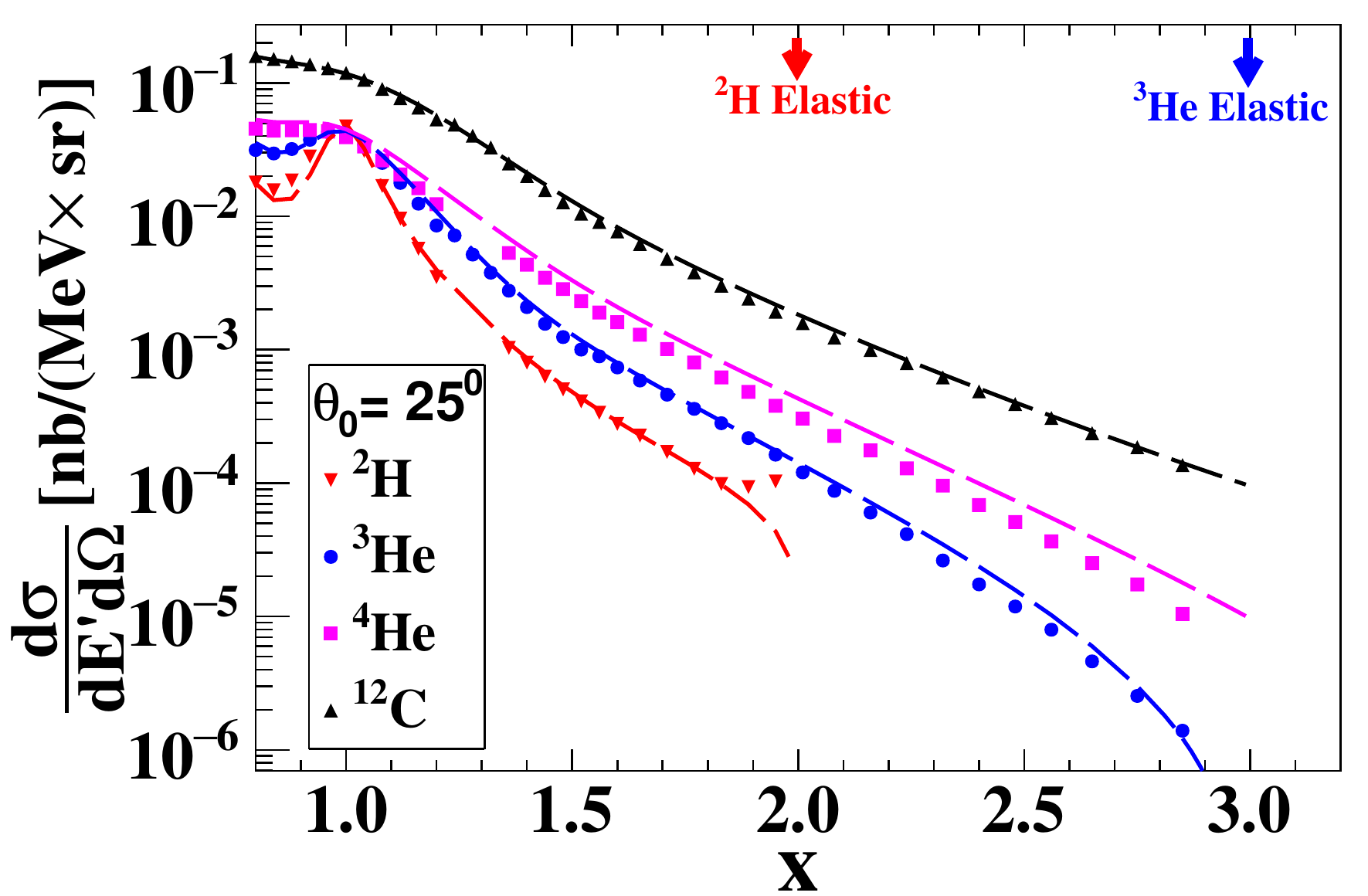}
    \end{center}
    \caption{(Color online) Cross sections of $^2$H, $^3$He, $^4$He and $^{12}$C at $25^{\circ}$. Only statistical
      uncertainties are shown. The dashed lines show our cross section model.}
        \label{xs}
\end{figure}

The inclusive cross sections for $^2$H $^{3}$He, $^{4}$He and $^{12}$C at a scattering angle of
$25^{\circ}$ are shown in Fig.~\ref{xs}. The $^3$He cross section falls more rapidly than the other nuclei
for $x>2.5$, generating the rise in the $^4$He/$^3$He ratios discussed above. In the naive SRC model, it is
assumed that the high-$x$ cross section comes from the contributions of \textit{stationary} 2N- and 3N-SRCs.
The prediction of scaling in this model breaks down due to the difference between stationary SRC in $^2$H
(or $^3$He) and moving SRCs in heavier nuclei. For the most recent extraction of 2N-SRCs from the A/$^2$H
ratios~\cite{fomin2012}, the effect of the 2N-SRC motion in heavier nuclei was estimated and found to give a
small enhancement of the ratio in the plateau region, with little distortion of the shape until $x >
1.9$~\cite{fomin2012} where the ratio increases rapidly to infinity.

\section{CONCLUSIONS}
We have performed high-statistics measurements of the $^4$He/$^3$He cross section ratio over a range of $Q^2$,
confirming the results of the low-statistics measurements from Hall C~\cite{fomin2012} and showing
a clear disagreement with the CLAS data~\cite{PhysRevLett.96.082501} for $x>2$. This supports the idea that
the large-$x$ CLAS data were limited by bin-migration effects due to the spectrometer's modest momentum
resolution~\cite{Higinbotham:2014xna}. We do not observe the plateau predicted by the naive SRC model, but
explain why the prediction for the ratios in the 3N-SRC regime are not as robust as those for
2N-SRC. While we do not observe the predicted plateau, this does not mean that 3N-SRCs are
unimportant in this region. Even if the cross section is dominated by 3N-SRCs, the inclusive scattering
ratios may not show a clean plateau due to the motion of 3N-SRCs in $A>3$ nuclei.

While our A/$^3$He ratios do not provide indication of 3N-SRCs,
they do provide important new measurements
of the cross section ratios at $x>1$ that can be used to test models of 2N and 3N-SRCs.
Further insight into the high-momentum components can be obtained by comparing the $^3$He cross section
at large $x$ with a model including one-body and 2N-SRC contributions, after accounting for the center-of-mass
motion of 2N-SRCs in $^3$He. A significant 3N-SRC contribution would increase the cross section relative to
a model without explicit 3N-SRC contributions. However, because this is a comparison to theory, rather than
a direct comparison of SRCs within two nuclei, one can no longer rely on final-state interactions canceling, and these
effects would have to be modeled.

Further measurements of this kind should provide improved sensitivity to 3N configurations in nuclei.
The biggest obstacle appears the modest $Q^2$ values of our new data and the limited region in $x$ where the correction for the motion of 3N-SRCs in heavy nuclei is small.
Additional JLab experiments are planned which will significantly extend the $Q^2$ range
for a variety of light and heavy nuclei~\cite{e1206105}, and make high-precision comparisons of scattering
from $^3$He and $^3$H~\cite{e1211112} to examine the isospin structure at larger momenta in nuclei with
very similar structure.


\begin{acknowledgments}

We acknowledge the outstanding support from the Jefferson Lab Hall A technical staff and the
JLab target group. This work was supported in part by the DOE Office of Science, Office of Nuclear Physics,
contract DE-AC05-06OR23177, under which JSA, LLC operates JLab, DOE contracts DE-AC02-06CH11357 and
DE-FG02-96ER40950, the National Science Foundation, and the UK Science and Technology
Facilities Council (ST/J000175/1,ST/G008604/1).

\end{acknowledgments}

\bibliographystyle{h-physrev3}
\bibliography{e08014}

\end{document}

%% file: authors.tex
\newcommand*{\JLAB}{Thomas Jefferson National Accelerator Facility, Newport News, VA 23606}
\newcommand*{\TLV}{Tel Aviv University, Tel Aviv 69978, Israel}
\newcommand*{\MIT}{Massachusetts Institute of Technology, Cambridge, MA 02139}
\newcommand*{\KENT}{Kent State University, Kent, OH 44242}
\newcommand*{\DOMINION}{Old Dominion University, Norfolk, VA 23529}
\newcommand*{\CALIF}{California State University, Los Angeles, Los Angeles, CA 90032}
\newcommand*{\Hampton}{Hampton University, Hampton, VA 23668}
\newcommand*{\PENNSYLVANIA}{Pennsylvania State University, State College, PA 16801}
\newcommand*{\Paris}{Institut de Physique Nucl\'{e}aire (UMR 8608), CNRS/IN2P3 - Universit\'e Paris-Sud, F-91406 Orsay Cedex, France}
\newcommand*{\Syracuse}{Syracuse University, Syracuse, NY 13244}
\newcommand*{\Kentucky}{University of Kentucky, Lexington, KY 40506}
\newcommand*{\William}{College of William and Mary, Williamsburg, VA 23187}
\newcommand*{\Virginia}{University of Virginia, Charlottesville, VA 22904}
\newcommand*{\Halifax}{Saint Mary's University, Halifax, Nova Scotia, Canada}
\newcommand*{\Glasgow}{University of Glasgow, Glasgow G12 8QQ, Scotland, United Kingdom}
\newcommand*{\Temple}{Temple University, Philadelphia, PA 19122}
\newcommand*{\Argonne}{Physics Division, Argonne National Laboratory, Argonne, IL 60439}
\newcommand*{\China}{China Institute of Atomic Energy, Beijing, China}
\newcommand*{\NRCN}{Nuclear Research Center Negev, Beer-Sheva, Israel}
\newcommand*{\Catania}{Universita di Catania, Catania, Italy}
\newcommand*{\Dequense}{Duquesne University, Pittsburgh, PA 15282}
\newcommand*{\Pittsburgh}{Carnegie Mellon University, Pittsburgh, PA 15213}
\newcommand*{\LongwoodUniv}{Longwood University, Farmville, VA 23909}
\newcommand*{\Florida}{Florida International University, Miami, FL 33199}
\newcommand*{\Tallahassee}{Florida State University, Tallahassee, FL 32306}
\newcommand*{\INFN}{INFN, Sezione Sanit\`{a} and Istituto Superiore di Sanit\`{a}, 00161 Rome, Italy}
\newcommand*{\INFNBari}{INFN, Sezione di Bari and University of Bari, I-70126 Bari, Italy}
\newcommand*{\Ohio}{Ohio University, Athens, OH 45701}
\newcommand*{\Tennessee}{University of Tennessee, Knoxville, TN 37996}
\newcommand*{\Kharkov}{Kharkov Institute of Physics and Technology, Kharkov 61108, Ukraine}
\newcommand*{\LOSALAMOS}{Los Alamos National Laboratory, Los Alamos, NM 87545}
\newcommand*{\Duke}{Duke University, Durham, NC 27708}
\newcommand*{\Texas}{University of Texas, Houston, TX 77030}
\newcommand*{\Seoul}{Seoul National University, Seoul, Korea}
\newcommand*{\Indiana}{Indiana University, Bloomington, IN 47405}
\newcommand*{\Hampshire}{University of New Hampshire, Durham, NH 03824}
\newcommand*{\Blacksburg}{Virginia Polytechnic Inst. and State Univ., Blacksburg, VA 24061}
\newcommand*{\France}{Universit\'e Blaise Pascal/IN2P3, F-63177 Aubi\`ere, France}
\newcommand*{\Mississippi}{Mississippi State University, Mississippi State, MS 39762}
\newcommand*{\Austin}{The University of Texas at Austin, Austin, Texas 78712}
\newcommand*{\Norfolk}{Norfolk State University, Norfolk, VA 23504}
\newcommand*{\Lanzhou}{Lanzhou University, Lanzhou, China}
\newcommand*{\Hebrew}{Racah Institute of Physics, Hebrew University of Jerusalem, Jerusalem, Israel}
\newcommand*{\Rutgers}{Rutgers, The State University of New Jersey, Piscataway, NJ 08855}
\newcommand*{\Yerevan}{Yerevan Physics Institute, Yerevan 375036, Armenia}
\newcommand*{\Ljubljana}{Faculty of Mathematics and Physics, University of Ljubljana, Ljubljana, Slovenia}
\newcommand*{\Michigan}{Northern Michigan University, Marquette, MI 49855}
\newcommand*{\Hefei}{University of Science and Technology, Hefei, China}	
\newcommand*{\Jozef}{Jozef Stefan Institute, Ljubljana, Slovenia}
\newcommand*{\Ecole}{CEA Saclay, F-91191 Gif-sur-Yvette, France}
\newcommand*{\Massachusetts}{University of Massachusetts, Amherst, MA 01006}

\author{Z. Ye}
\affiliation{\Argonne}
\affiliation{\Virginia}
\affiliation{\Duke}
\author{P. Solvignon}
\thanks{deceased}
\affiliation{\Hampshire}
\affiliation{\JLAB}
\author{D. Nguyen}
\affiliation{\Virginia}
\author{P. Aguilera}
\affiliation{\Paris}
\author{Z. Ahmed}
\affiliation{\Syracuse}
\author{H. Albataineh}
\affiliation{\DOMINION}
\author{K. Allada}
\affiliation{\JLAB}
\author{B. Anderson}
\affiliation{\KENT}
\author{D. Anez}
\affiliation{\Halifax}	
\author{K. Aniol}
\affiliation{\CALIF}
% ...

\collaboration{The Jefferson Lab Hall A Collaboration}